\documentstyle[psfig,amssymb]{mn}
\title[Searching for X-ray selected normal
galaxies]{The Needles in the Haystack survey: searching for X-ray
selected normal galaxies} 
 
\author[Georgakakis et al.] {A. E. Georgakakis$^{1}$
  I. Georgantopoulos$^{1}$, S. Basilakos$^{1}$, M. Plionis$^{1,2}$
  \\ \\
  $^1$ Institute of Astronomy \& Astrophysics, National Observatory of
  Athens, I. Metaxa \& V. Pavlou, Athens, 15236, Greece \\ 
  $^{2}$ Instituto Nacional de Astrof\'isica \'Optica y
    Electr\'onica, AP 51 y 216, 72000, Puebla, Pue, M\'exico
}
\begin{document}
\maketitle  

\begin{abstract}
In this paper we present the first results from an on-going
serendipitous  survey aiming to identify  X-ray selected `normal'
galaxies (i.e. not AGN dominated) by combining archival XMM-{\it
Newton} data  with the Sloan Digital Sky Survey. In the first $\rm 4.5\, 
deg^2$ of this program we have identified a total of 11 `normal'
galaxy candidates (8 of them with optical spectroscopy) with fluxes
$f_X(\rm 0.5 - 8 \, keV) \approx 10^{-15} - 10^{-13} \, erg \, s^{-1}
\, cm^{-2}$. These sources are selected to have low X-ray--to--optical
flux ratio ($\log f_X/f_{opt}   \la -2$), soft X-ray spectral
properties and optical spectra, when available, consistent with the
presence of a  stellar ionising continuum.  These sources comprise
both early and late type systems at  redshifts $z\la0.2$ with
luminosities $L_X(\rm 0.5 - 8\, keV) \approx 10^{39} - 10^{42}\, erg\,
s^{-1}$. This dataset provides the first tight constraint on the
surface density of X-ray selected `normal' galaxies at relatively
bright fluxes spanning two orders of magnitude ($\rm 10^{-15} -
10^{-13} \, erg \, s^{-1} \, cm^{-2}$). The slope of the `normal'
galaxy $\log N - \log S$ in the above flux range is estimated
$-1.4\pm0.3$ consistent with the euclidean prediction. We also
discuss the prospects of `normal' galaxy studies at X-ray wavelengths
using both our continuously expanding survey and future X-ray
missions.    
\end{abstract}


\section{Introduction}\label{sec_intro}
In the local Universe ($\la100$\,Mpc) `normal' galaxies  with X-ray
emission dominated by stellar processes have been extensively studied
using data from targeted observations of selected systems (e.g. Read,
Ponman \&  Strickland 1997; Shapley, Fabbiano \& Eskridge 2001; Zezas
et al. 2003; Fabbiano et al. 2003; Jeltema et al. 2003).  Although
these individual observations provide valuable information on the
nature and the X-ray properties of `normal' galaxies they cannot  be
used for statistical studies (e.g. $\log N - \log S$, luminosity
function, clustering properties).  

It is only recently that ultra-deep {\it Chandra} surveys reaching
fluxes  $f(\rm 0.5 - 2 \,keV) \approx 3\times10^{-17} \, erg\,
s^{-1}\, cm^{-2}$ (e.g. Brandt et al. 2001; Giacconi et al. 2002;
Alexander et al. 2003) have
provided the first X-ray selected sample of `normal' galaxy candidates
(Hornschemeier et al. 2003; Norman et al. 2004). These
objects although too faint to significantly contribute to the X-ray
background (XRB) are detected in increasing numbers with decreasing
flux and are likely to outnumber powerful AGNs below $f(\rm 0.5 - 2
\,keV) \approx 10^{-17} \, erg\, s^{-1}\, cm^{-2}$ (Hornschemeier et
al. 2003).   

The studies above however, concentrate on `normal' galaxies at faint 
X-ray limits ($\approx 10^{-16} \rm \, erg \, s^{-1} \, 
cm^{-2}$) and moderate redshifts $z=0.1-1$ (median $\approx0.3$). At
brighter fluxes ($\approx 10^{-15} - 10^{-13} \rm \, erg \, s^{-1} \,
cm^{-2}$)  probing, on average,  lower redshifts, $z\la0.2$, our
knowledge on these  systems remains scanty. This is because `normal'
galaxies at bright X-ray fluxes have low  sky density.  Therefore,
wide area surveys are required to compile large enough  samples to
explore their statistical properties (luminosity function, surface
density; Georgakakis et al. 2004). Such  a low-$z$ study is essential
to interpret the X-ray `normal' galaxies in deeper  X-ray surveys and
to investigate their evolution. This is particularly important since
the X-ray  emission is the only observable window available that
allows one to probe the evolution of the X-ray binaries and the hot gas
in galaxies. This provides information on the star-formation history
of the Universe that is complementary to that obtained from other 
wavelengths (Ghosh \& White 2001; Hornschemeier et al. 2002;
Georgakakis et al. 2003; Norman et al. 2004).    

Wide areal coverage is difficult to achieve with the {\it Chandra}
observatory due its limited field--of--view. On the contrary, XMM-{\it
Newton} with 4 times larger field-of-view provides an ideal platform
for such a study. In this paper we exploit the capabilities and the
large volume of archival data of the XMM-{\it  Newton} to
serendipitously identify  `normal' galaxy candidates in public fields
selected to overlap with the Sloan Digital Sky Survey (SDSS; York et
al. 2000) to exploit the superb and uniform 5-band optical photometry
and spectroscopy available in this area. In this paper we report the
first results of this survey (Needles in the Haystack Survey; NHS)
using a pilot  sample covering a total area of $\rm \approx 4.5\,deg^2$
in the flux range $f_X(\rm 0.5 - 8 \, keV) \approx 10^{-15} -
10^{-13}\, erg \, s^{-1}\, cm^{-2}$. This dataset is continuously
expanded as more SDSS fields and XMM-{\it Newton} archival data become
available. Throughout this paper we adopt $\rm H_{o} = 65 \, km \,
s^{-1} \, Mpc^{-1}$, $\rm \Omega_{M} = 0.3$ and $\rm  \Omega_{\Lambda}
= 0.7$.      

\section{The sample}\label{sec_survey}

In this paper we use XMM-{\it Newton} archival observations, with a
proprietary period that expired before September 2003, that overlap
with first data release of the SDSS (DR1; Stoughton et al. 2002). Only
observations that use the EPIC (European Photon Imaging Camera; 
Str\"uder et al. 2001; Turner et al. 2001) cameras as the prime
instrument operated in full frame mode were employed. For fields
observed more than once with the XMM-{\it Newton} we use the deeper of
the multiple observations. A total of 20 fields are used with Galactic
$\rm N_H$ in the range $1.4-13.2\times10^{20}\rm \, cm^{-2}$ (mean
$3.7 \times 10^{20}\rm \, cm^{-2}$) and PN good time intervals between
2 and 67\,ks (mean 18.5\,ks). These observations are complemented with
8 XMM-{\it Newton}  pointings from the northern region of the wide
area, shallow XMM-{\it Newton}/2dF survey that also overlaps with the
SDSS (Georgakakis et al. 2004).  

\begin{figure}
\centerline{\psfig{figure=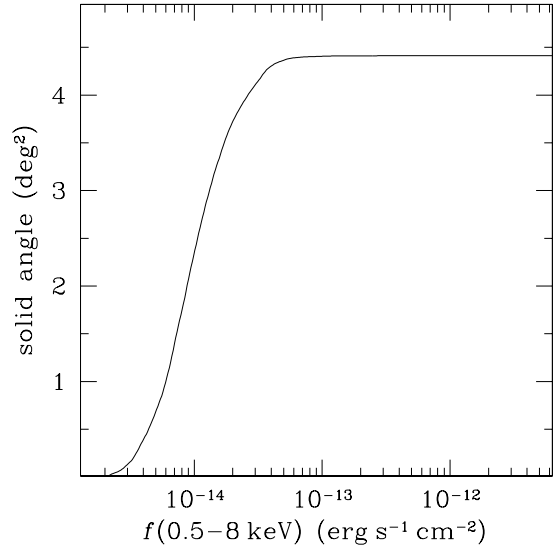,width=3in,angle=0}}
\caption
 {Solid angle as a function of limiting flux desnity ($5\sigma$) in the
total 0.5-8\,keV band for our survey. 
 }\label{area}
\end{figure}

The X-ray data reduction, source detection and flux estimation are
carried out using methods described in Georgakakis et
al. (2004). Source extraction is performed in the 0.5-8\,keV merged
PN+MOS images using the {\sc ewavelet} task of {\sc sas} with a
detection threshold of $5\sigma$. The extracted sources for each field
were visually inspected and spurious detections clearly associated
with CCD gaps, hot pixels or lying close to the edge of the field of
view were removed.  Figure \ref{area} shows the solid angle covered 
by our survey as a function of the  0.5-8\,keV limiting flux. 

To optically identify the X-ray sources we follow the method of Downes
et al. (1986) as described in Georgakakis et al. (2004) to calculate
the probability, $P$,  a given candidate is the true
identification. Here we apply an upper limit in the search radius,
$r<7\rm\,arcsec$ and a cutoff in the probability, $P<0.05$, to limit
the optical  identifications to those candidates that are least likely
to be chance coincidences. 

`Normal' galaxy candidates are selected to have  (i) extended optical
light profile (i.e. resolved) to avoid contamination of the sample by
Galactic stars, (ii) X-ray--to--optical flux ratio $\log f_X / f_{opt}
< -2$, two orders of magnitude lower than typical AGNs (for the
definition of $\log f_X / f_{opt}$ see Georgakakis et al. 2004). The
sample of `normal' galaxy candidates is presented in Table \ref{tab2}.  

The archival X-ray data used here include targeted  
observations of nearby normal galaxies with low
X-ray--to--optical flux ratio. Such sources have been excluded from
Table \ref{tab2}. Moreover, a number of `normal' galaxy  candidates
although not the prime target of the XMM-{\it Newton} pointing lie at
the same redshift as the prime target and are therefore most likely
directly associated with it (e.g. cluster or group members). These
sources are marked in Table \ref{tab2}  and are excluded from any
statistical analysis. 

\section{Optical and X-ray properties}\label{sec_sample}

Low luminosity Seyferts also lie in the $\log f_X /
f_{opt}\la-1$  regime and may contaminate our `normal' galaxy sample
(Lehmann et al. 2001). We explore  this possibility using both the
X-ray and the optical spectroscopic information available for our
sources.  

For sources \#2, 4, 6, 12 and 15 in Table \ref{tab2} we can constrain  the 
nature of the central engine using the diagnostic diagrams proposed
by Veilleux \& Osterbrock (1987) that employ intensity line ratios
that are least sensitive to dust reddening (H$\beta$/[O\,III] 5007,
[S\,II]  6716+31/H$\alpha$, [O\,I]  6300/H$\alpha$ and [N\,II] 
6583/H$\alpha$). We adopt the classification scheme of Ho, Fillipenko
\& Sargent (1997) and find that sources \#2, 6, 12 and 15 have H\,II
type spectra while \#4 is classified as a Seyfert 2. As will be
discussed below the classification of source \#4 is  consistent with
its absorbed X-ray spectrum.  

For the remaining sources in Table \ref{tab2} the classification using 
diagnostic diagrams cannot be applied because either they have
absorption lines only (\#9, 10, 16, 17), no spectroscopic
information (\#1, 5, 7, 11, 14), or both the $\rm H\beta$ and
the $\rm   [ OIII ] \, 5007\,\AA$ lines, essential for spectral
classification, are absent from the optical spectrum (\#3, 8,
13). The latter sources have strong absorption features from a
dominant evolved stellar population suggesting  early and intermediate
Hubble  types. Galaxies with both absorption and emission lines have
been extensively studied by Dressler et al. (1999) and Poggianti et
al. (2000) on the basis of the equivalent widths of the H$\delta$
(absorption) and  $\rm [OII]\,3727\AA$ (emission) lines. Our sources
have EW$\approx -1$ (absorption) and $\rm \approx10 \,\AA$ (emission)
for the H$\delta$ and $\rm [OII]\,3727\AA$ lines
respectively. Following the classification of Dressler et al. (1999)
the evidence above suggests e(c)-type galaxies which are believed to
be systems with constant star-formation rate over a period of time
resulting in a mix of young and evolved stellar populations. For these
sources however, we cannot exclude the possibility of low-luminosity
AGN activity diluted by the evolved stellar population
(e.g. Severgnini et al. 2003)  

We can further constrain the nature of our `normal' galaxy candidates
via X-ray spectral analysis using  the {\sc xspec} v11.2  
package. We use the
C-statistic technique (Cash 1979) specifically developed to extract
information from even low signal-to-noise ratio spectra. The data are
grouped to have at least one count per bin. An absorbed power-law
model is fit to the data fixing  the power law index  to $\Gamma=1.9$
and leaving $\rm N_H$ free parameter. The results are shown in Table
\ref{tab2}. The majority of the `normal' galaxy candidates have soft
X-ray spectral properties suggesting obscuration consistent with the
Galactic $\rm N_H$.  The only exception is source \#4 which is
significantly absorbed. 

The final sample of `normal' galaxy candidates that are not associated
with the prime target of a given XMM-{\it Newton} pointing and not
showing evidence (e.g. X-ray or optical spectra) for obscured AGN
activity comprises 11 sources,  \#1, 2, 3, 5, 6, 8, 9, 10, 11, 13, 15 in
Table \ref{tab2}. Of these 11 sources only sources \#1, 5 and 11 do
not have optical spectroscopy available.

\begin{table*} 
\footnotesize 
\begin{center} 
\begin{tabular}{l ccc ccc  cc cc}
\hline 
ID &
$\alpha_X$ & 
$\delta_X$ & 
$P$    &
$r$    &
$z$    &
net    &
$f_X(\rm 0.5-8\,keV)$ &
$L_X(\rm 0.5-8\,keV)$ &
%$\Gamma$ &
$\rm N_H$ &
type$^a$ \\
 
 &
(J2000)&
(J2000)&
(per cent) &
(mag)&
     &
counts &
($10^{-14}\, \rm erg \, s^{-1} \, cm^{-2}$)  &
($10^{41} \,\rm erg \, s^{-1}$)  & 
%    &
($\rm 10^{22}\,cm^{-2}$) &
    \\
\hline

% source 18, HAWAII167 field
1$^b$  & 03 58 05.22 & +01 09 48.96 & 0.01 & 14.93 & 0.074 
   & 53  
   & $1.52\pm0.34$ & 2.36
   %& $3.40^{+0.87}_{-0.55}$ 
   & $<0.04$ & early \\

% source 5, this is the IRAS detected starburst
2 & 08 30 59.21 &  +52 37 47.75 & 0.80 & 17.69 & 0.1363 
  & 48
  & $0.34\pm0.13$ & 2.01
%  &  $2.45^{+1.25}_{-1.15}$ 
  & $<0.13$   & NL \\
 
% source 8, this is EA source
3 & 08 31 14.54 & +52 42 24.77 & 0.01 & 15.24 & 0.0641 
  & 120
  & $0.56\pm0.10$ & 0.66
%  & $2.39^{+0.28}_{-0.30}$ 
  & $<0.02$ & EA \\

% source 7 OBSCURED AGN
4 & 08 31 39.09 & +52 42 05.51 & 0.02 & 15.70 & 0.0587 
  & 337
  & $2.07\pm0.14$ & 2.00
%  & $-0.04^{+0.04}_{-0.26}$ 
  & $14.9^{+0.9}_{-2.6}$ & NL \\
\\
% source 9
5$^b$  & 08 32 02.58 & +52 47 13.74 & 0.01 & 16.01 & 0.105 
  & 55
  & $0.25\pm0.09$ & 0.81
%  & $2.07^{+0.44}_{-0.40}$ 
  & $<0.1$ & late \\

% source 4 , this is the IRAS detected starburst
6 & 08 32 28.14 &  +52 36 22.36 & $<$0.01 & 13.82 & 0.0170 
  & 716
  & $6.43\pm0.29$ & 0.52
%  & $2.50^{+0.08}_{-0.15}$ 
  & $<0.02$ & NL \\

%source 3 photometric-z same as target-z, ABELL\,773
7$^{b,c}$  & 09 17 58.59 & +51 51 04.75 & 0.31 & 16.66 & 0.219 
  & 22
  & $0.61\pm0.26$ & 9.76
%  & $2.4^{+0.7}_{-0.6}$ 
  & $<0.06$ & early \\

% source 12
8 & 09 35 18.39 & +61 28 31.58 & 0.27 & 16.50 & 0.1241 
   & 54
   & $0.96\pm0.25$ & 4.48
%   & $2.04^{+0.66}_{-0.47}$ 
   & $<0.08$ & EA \\
\\   
% source 13
9 & 09 36 19.46 & +61 27 23.11 & 0.06 & 16.55 & 0.1311 
  & 67
  & $0.95\pm0.21$ & 5.25
  % & $2.72^{+0.44}_{-0.42}$ 
   & $<0.03$ & AB \\

% source 10 this is  NGC\,4510
10 & 12 31 47.21 & +64 14 01.68 & $<0.01$ & 13.01 & 0.009 
   & 67
   & $1.95\pm0.14$ & 0.047
%   & $2.02^{+0.25}_{-0.19}$ 
   & $<0.02$ & AB \\

% source 11
11$^b$  & 12 32 53.15 & +64 08 50.93 & 0.17 & 15.43 &  0.139 
   & 50
   & $1.03\pm0.26$ &  6.12
%   & $2.48^{+0.68}_{-0.52}$ 
   & $<0.06$ & late \\

% source 1 same as target-z, NGC\,4666
12$^{c}$ & 12 45 31.99 & --00 32 08.59 & 0.01 & 13.12 & 0.0055 
  & 353
  & $2.83\pm0.19$ & 0.02 
%  & $1.42^{+0.24}_{-0.09}$ 
  & $0.11^{+0.07}_{-0.03}$ &  NL \\
\\
% source 2
13 & 12 44 52.26 & --00 25 49.66 & 0.01 & 15.34 & 0.0823 
  & 95
  & $0.57\pm0.12$ & 1.10 
%  & $2.72^{+0.44}_{-0.42}$ 
  &  $<0.03$ & EA \\

% source 15 photo-z same as target-z, ABELL\,1674
14$^{b,c}$  & 13 03 01.13 & +67 25 17.76 & 0.16 & 16.16 &  0.109 
   & 49
   & $1.38\pm0.35$ & 4.90
   %& $2.30^{+0.40}_{-0.40}$ 
   & $0.01_{+0.07}^{-0.01}$ & early \\

% UM 603
15 &  13 41 37.85 & --00 25 55.30  & 0.05 & 16.86 &  0.053
   & 30
   & $1.95\pm0.52$ & 1.44
   %& $2.14^{+0.70}_{-0.46}$ 
   & $<0.09$ & NL \\

% source 16 same as target-z, ABELL 2670
16$^{c}$ & 23 53 40.52 &--10 24 19.91 & 0.02 & 15.07 & 0.0741 
   & 52
   & $2.52\pm0.52$ & 3.92
   %& $2.60^{+0.30}_{-0.30}$ 
   & $<0.03$ & AB \\

% source 17 same as target-z, ABELL 2670
17$^{c}$ & 23 54 05.69 &--10 18 30.64 & 0.06 & 15.70 & 0.0734 
   & 50
   & $1.06\pm0.33$ & 1.61
   %& $2.40^{+0.20}_{-0.20}$ 
   & $<0.02$ & AB \\

\hline
%\multicolumn{8}{l}{$^1$: 0.5-8\,keV X-ray flux in units of $10^{-14}\,
%erg \, s^{-1} \, cm^{-2}$}\\
%\multicolumn{8}{l}{$^2$: 0.5-8\,keV X-ray luminosity in units of
%$10^{41}\, erg\, s^{-1}\, cm^{-2}$}\\
\multicolumn{11}{l}{$^a$AB: absorption lines; NL: Narrow emission
lines: EA: both narrow emission and absorption lines} \\
\multicolumn{11}{l}{$^b$photometric redshift and classification from
the SDSS}\\
\multicolumn{11}{l}{$^c$source at the same redshift as the target of
the XMM-{\it Newton} pointing.}\\

\multicolumn{11}{l}{The columns are: 1: identification number; 2: right
ascension of the X-ray source; 3: declination of the X-ray source;}\\

\multicolumn{11}{l}{4: probability the optical counterpart is a
spurious alignment; 5: optical magnitude; 6: spectroscopic or 
photometric redshift from the SDSS;}\\

\multicolumn{11}{l}{7: 0.5-8\,keV net counts; 8: X-ray flux in the 0.5-8\,keV spectral band in
units of $10^{-14} \, \rm erg \, s^{-1} \, cm^{-2}$;}\\

\multicolumn{11}{l}{9: 0.5-8\,keV
X-ray  luminosity in units of $10^{41}\, \rm erg \, s^{-1}$;10: 
$\rm N_H$ from the X-ray spectrum assuming 
$\Gamma=1.9$; 11: spectral type.}\\
\end{tabular} 
\end{center} 
\caption{
The candidate `normal' galaxy sample. 
%The columns are: 1:
%identification number; 2: right ascension of the X-ray source; 3:
%declination of the X-ray source; 4: probability the optical
%counterpart is a spurious alignment; 5: optical $r$-band magnitude; 6:
%spectroscopic or photometric redshift $z$ from the SDSS; 7: X-ray flux
%in the 0.5-8\,keV spectral band in units of $10^{-14} \, \rm erg \, s^{-1} 
%\, cm^{-2}$; 7: 0.5-8\,keV X-ray  luminosity in units of $10^{41}\, \rm erg \,
%s^{-1}$;  8: hardness ratio, HR,  estimated from the count rates in
%the 0.5-2 and 2-8\,keV spectral bands; 9: spectral type using either
%diagnostic emission line ratios or the best fit spectral energy
%distribution in the case of photometric redshifts.
}\label{tab2} 
\normalsize  
\end{table*}

%\begin{table*}
%\footnotesize
%\begin{center}
%\begin{tabular}{l ccc ccc}
%
%\hline
%
%ID & \multicolumn{3}{c}{power-law model} & \multicolumn{3}{c}{Raymond-Smith model} \\
%   & $\Gamma$ & N$_H$  & $\chi^2$/d.o.f        &  kT & N$_H$  & $\chi^2$/d.o.f \\
%   &          & ($\rm 10^{22}\,cm^{-2}$) &       &  (K) & ($\rm 10^{22}\,cm^{-2}$)  &  \\
%\hline
%1  &  $1.38^{+0.41}_{-0.21}$  & $<0.08$ & 35.0/29 &
%      $14.8^{+49.2}_{-8.6}$ & 0.02(fixed) & 34.7/28 \\ 
%
%4  &  $6.0^{+1.0}_{-0.8}$    & $0.61^{+0.13}_{-0.11}$  & 78.6/55    &
%      $0.7^{+0.1}_{-0.1}$    & $0.11^{+0.05}_{-0.04}$  & 45.0/54    \\
%
%6  &  -- & -- & -- & 
%      -- & -- & -- \\
%
%7  &  $2.2^{+0.6}_{-0.4}$    & $<0.07$  & 21.2/14    &
%      $2.0^{+2.1}_{-0.3}$    &$<0.04$   & 25.5/14    \\
%
%\hline
%\end{tabular}
%\end{center}
%\caption{Comparison between power-law and Raymond-Smith models 
%for sources with sufficient X-ray photon statistics to perform $chi^2$
%spectral analysis.  
%}\label{tab3}
%\normalsize
%\end{table*}

\begin{figure}
\centerline{\psfig{figure=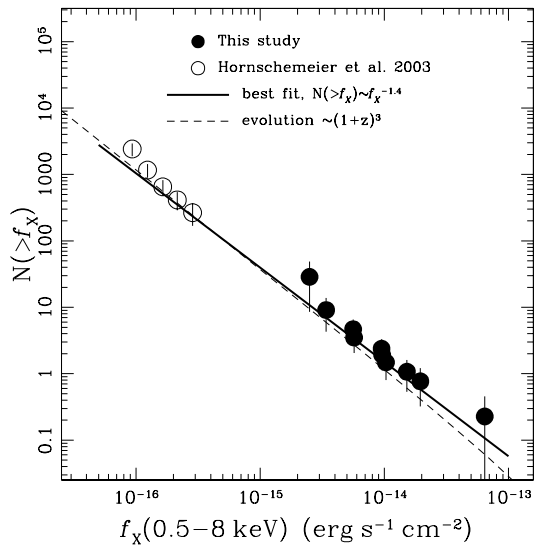,width=3in,angle=0}}
\caption
 {Cumulative `normal' galaxy counts. Filled circles are the `normal'
 galaxy candidates in the present study. We plot 10 instead of 11
 points because sources \#10 and 15 have the same flux and are plotted
 as one point. The open circles are the source counts of Hornschemeier
 et al. (2003). The continuous line is the best fit to the $\log N -
 \log S$ at bright fluxes from the present sample. The dashed line is
 the model prediction of the Georgantopoulos et al. (1999) X-ray
 luminosity function of H\,II galaxies assuming evolution of the form
 $L_X\sim(1+z)^{3}$.   
 }\label{fig_logn}
\end{figure}

\begin{figure}
\centerline{\psfig{figure=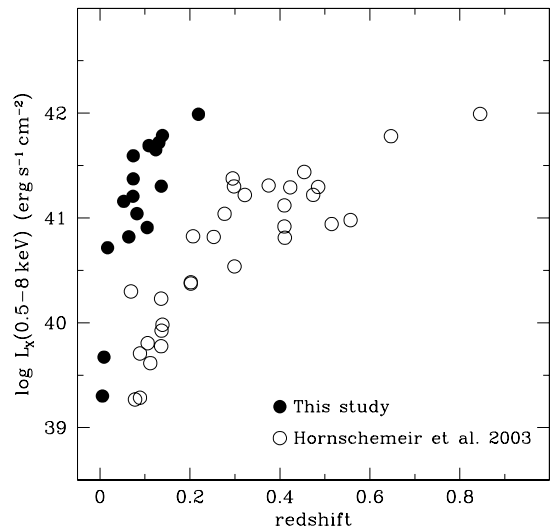,width=3in,angle=0}}
\caption
 {$L_X(\rm 0.5 - 8\,keV)$ against redshift. Filled circles are for our
 sample of `normal' galaxy candidates (including sources associated
 with the prime target of a given XMM-{\it Newton} pointing). Open
 circles represent the sample of Hornschemeier et al. (2003).  The two
 surveys cover different but complementary regions of the $L_X - z$
 space.   
 }\label{fig_lxz}
\end{figure}

\section{Discussion}\label{discussion}
In this paper we report on an on-going program aiming to identify
X-ray selected `normal' galaxies in the flux range $f(\rm
0.5-8\,keV)\approx10^{-15}-10^{-13}\,erg\,s^{-1}\,cm^{-2}$ by
combining archival XMM-{\it Newton} data with the SDSS. At present our
on-going survey covers an area of $\rm 4.5\deg^2$ and has identified a
total of 11 `normal' galaxy candidates with $\log f_X /
f_{opt}<-2$. Unlike deeper samples our sources are relatively bright
at both the optical and the X-ray wavelengths allowing detailed study
of their nature thus, minimising any AGN contamination. Both the soft
X-ray spectra and the optical spectroscopic properties are consistent
with  the absence of nuclear  activity.

Figure 2 plots the $\log N - \log S$ of our sample in comparison with
that of `normal' galaxy candidates identified in the {\it Chandra}
Deep Field North  survey probing fluxes  $\rm \approx   10^{-16}\,erg
\,s^{-1} \,cm^{-2}$ (Hornschemeier et al. 2003).  The present study
provides the only constraints on the number density of `normal'
galaxies at relatively bright fluxes in the range $f(\rm  0.5 - 8 \,
keV) \approx 10^{-15} - 10^{-13} \,erg \, s^{-1} \, cm^{-2}$ spanning   
two orders of magnitude. The differential `normal' galaxy counts in the
0.5-8\,keV band are fit by a power law  yielding a slope of
$-1.4\pm0.3$ for the cumulative number counts. This is in excellent
agreement with the results of  Hornschemeier et  al. (2003) at fainter
fluxes ($-1.46^{+0.28}_{-0.30}$). Figure 2 also shows that our  wide
area medium-deep serendipitous survey is complementary to the
ultra-deep pencil beam  {\it Chandra} observations. This is further
demonstrated in Figure 3 plotting the $L - z$  plane for both the  
present sample and that of  Hornschemeier et  al. (2003). The two
surveys cover different but complementary regions of the parameter
space with our sample probing the low--$z$, high--$L_X$  regime. Such a
wide $L - z$ plane coverage is essential to explore the evolution
of `normal' galaxies at X-ray wavelengths. Although our sample is not
large enough yet for such a study, we plan to expand it to $\rm
\approx 20\,deg^2$ by exploiting the 2nd data  release of the
SDSS. Our enlarged sample will comprise $\approx 40$ `normal' galaxies 
sufficient for luminosity function estimates providing an anchor point
at low redshifts, essential for comparison with higher-$z$ studies
(e.g. Hornschemeier et al. 2003; Georgakakis et al. 2003; Norman et
al. 2004).

However, even the deepest X-ray samples todate find `normal' galaxies
at relatively low-$z$ (median redshift $z\approx0.3$; Hornschemeier et
al. 2003) significantly lower than the star-formation peak at
$z\approx1-2$. This underlines the need for deeper
X-ray observations to use X-ray selected `normal' galaxies as 
cosmological probes. The ESA {\it XEUS} mission with its unparalleled
sensitivity will provide a great leap forward in these
studies. Assuming the goal  resolution of about 2\,arcsec FWHM for its
mirrors,  {\it XEUS} will  attain, in just 20\,ks, about equal
sensitivity to the deepest (2\,Ms) {\it Chandra} exposure. Extrapolation
of the $\log N- \log S$ relation derived here to $f_X (\rm 0.5 - 8
)=10^{-16} \, erg \, cm^{-2} \, s^{-1}$ (i.e. approximate flux limit
of the 2\,Ms {\it Chandra} Deep Field North; Alexander et al. 2003),
suggests that about 10 galaxies will be detected in the $25\, \rm
arcmin^2$ {\it XEUS} Wide Field Imager. Therefore, a few tens of
typical exposures with {\it  XEUS} will readily accumulate a large
sample of X-ray selected  galaxies for the study of the galaxy
luminosity function and evolution  at redshifts $z\la1$ with a median
redshift of $z\sim0.3$. Deeper exposures are necessary to probe the
star-formation peak at higher redshifts. In a deep 200\,ks exposure,
{\it XEUS} is   expected to probe fluxes as faint as $f_X(\rm 0.5 -
8\, keV)  \approx 5\times 10^{-18}\, erg \,s^{-1}\,cm^{-2}$ and to
detect more than 500 galaxies (an estimate  obtained by extrapolating
our $\log N - \log S$). At these flux levels the mean redshifts probed
are $z\sim 1$ as derived from the X-ray luminosity function of
Georgantopoulos et al. (1999) assuming evolution of the form
$L_X\sim(1+z)^3$. Therefore, combination of the   local XMM-{\it
Newton} galaxy samples with the {\it XEUS} data  will allow an
unprecedented study of the star-formation history at X-ray
wavelengths.

%\section{Acknowledgments}
% This work is jointly funded by the European Union  and the Greek
% Government  in the framework of the programme ``Promotion of
% Excellence in Technological Development and Research'', project
% ``X-ray Astrophysics with ESA's mission XMM''.   
%
%
% Funding for the creation and distribution of the SDSS Archive has
% been provided by the Alfred P. Sloan Foundation, the Participating
% Institutions, the National Aeronautics and Space Administration, the
% National Science Foundation, the U.S. Department of Energy, the
% Japanese Monbukagakusho, and the Max Planck Society. The SDSS Web
% site is http://www.sdss.org/. The SDSS is managed by the
% Astrophysical Research Consortium (ARC) for the Participating
% Institutions. The Participating Institutions are The University of
% Chicago, Fermilab, the Institute for Advanced Study, the Japan
% Participation Group, The Johns Hopkins University, Los Alamos
% National Laboratory, the Max-Planck-Institute for Astronomy (MPIA),
% the Max-Planck-Institute for Astrophysics (MPA), New Mexico State
% University, University of Pittsburgh, Princeton University, the
% United States Naval Observatory, and the University of Washington. 

\end{document}